\documentstyle[12pt]{article}

\newcommand{\resection}[1]{\setcounter{equation}{0}\section{#1}}

\newcommand{\EQ}{\begin{equation}}
\newcommand{\EN}{\end{equation}}
\newcommand{\bea}{\begin{eqnarray}}
\newcommand{\eea}{\end{eqnarray}}
\newcommand{\hs}{\hspace{0.1cm}}

\newcommand{\be}{\beta}
\newcommand{\zb}{\bar{z}}

\begin{document}
\setcounter{page}{0}
\topmargin 0pt
\oddsidemargin 5mm
\renewcommand{\thefootnote}{\fnsymbol{footnote}}
\newpage
\setcounter{page}{0}
\begin{titlepage}
\begin{flushright}
ISAS/EP/92/146 \\
Imperial/TP/91-92/31
\end{flushright}
\vspace{0.5cm}
\begin{center}
{\large {\bf Form Factors for Integrable Lagrangian Field Theories,
              the Sinh-Gordon Model}}

\vspace{1.8cm}
{\large A. Fring}\\
\vspace{0.5cm}
{\em The Blackett Laboratory, \\
Imperial College, London SW7 2BZ UK}\\
\vspace{1cm}
{\large G. Mussardo, P. Simonetti}\\
\vspace{0.5cm}
{\em International School for Advanced Studies,\\
34014 Trieste, Italy}\\

\end{center}
\vspace{1.2cm}

\renewcommand{\thefootnote}{\arabic{footnote}}
\setcounter{footnote}{0}

\begin{abstract}
{Using Watson's and the recursive equations satisfied by matrix
elements of local operators in two-dimensional integrable models, we
compute the form factors of the elementary field $\phi(x)$ and the
stress-energy tensor $T_{\mu\nu}(x)$ of Sinh-Gordon theory. Form
factors of operators with higher spin or with different asymptotic
behaviour can easily be deduced from them. The value of the correlation
functions are saturated by the form factors with lowest number of particle
terms. This is illustrated by an application of the form factors of the
trace of $T_{\mu\nu}(x)$ to the sum rule of the $c$-theorem.}
\end{abstract}
%\vspace{.3cm}
%\centerline{July 1992}
\end{titlepage}

\newpage

\resection{Introduction}

\,

Recent investigations on two-dimensional quantum field theories have
established the exact integrability for a variety of physically interesting
models with massive excitations. A rather simple characterization
of such theories may be given in terms of their scattering data, i.e.
their properties on mass-shell. In fact, the existence of an infinite number
of commuting conserved charges implies that the scattering processes which
occur in these theories preserve the number of particles and the set of their
asymptotic momenta \cite{ZZ}. The computation of the exact factorized
$S$-matrix may be performed by combining the standard requirements of
unitarity and crossing symmetry together with the symmetry properties of the
model [1-7]. In many cases, the conjectured $S$-matrix may be supported by
perturbative checks \cite{ZZ,AFZ,MC,Braden}.

The knowledge of the exact $S$-matrix can then be used to compute off-shell
quantities, like correlation functions of elementary or composite fields
of the integrable models under investigation. This can be achieved by
considering the form factors of local fields, which are matrix elements of
operators between asymptotic states. General properties of unitarity,
analyticity and locality lead to a system of functional equations for these
matrix elements which permit in many cases their explicit determination
[10-17]. The correlation functions are then written in terms of an infinite
sum over the multi-particle form factors.

In this paper we investigate one of the simplest integrable Lagrangian system,
namely the Sinh-Gordon theory. Many properties of this model are well
established, the exact $S$-matrix for instance was obtained in \cite{AFZ},
whereas the quantization of this theory has been studied in \cite{Sklyanin}.
Our objective is to derive expressions for the form factors of the elementary
field $\phi(x)$ and the energy-momentum tensor $T_{\mu\nu}(x)$, which are the
most representative operators of the odd and even sector of the $Z_2$
symmetry of this model.

The paper is organized as follows: in section 2 we discuss general properties
of form factors for integrable models, i.e. their analytic structure,
Watson's and the recursive equations which they satisfy. In section
3 we recall the basic properties of the Sinh-Gordon theory. Section 4 is
devoted to the explicit computation of form factors for this theory.
In section 5 we investigate the natural grading introduced in the
space of the form factors by the arbitrariness inherent Watson's equations
and, in particular, we show how form factors of operators with higher spin
can be obtained from the ones for $\phi(x)$ and $T_{\mu\nu}(x)$. In section 6
we make use of the form factors of the stress-energy tensor in order to
illustrate the $c$-theorem. In the last section there are our conclusions.

\resection{General Properties of Form Factors}

\,

Essential input for the computation of form factors is the knowledge of the
scattering matrix $S$. For two-dimensional integrable systems the expression
of the $S$-matrix is particularly simple and may be obtained explicitly for
several systems [1-7]. Since the dynamics is governed by an infinite number
of higher conservation laws, the scattering processes for integrable models are
purely elastic and the general n-particle S-matrix defined by
\EQ
S_n(p_1,\ldots,p_n)\,=\,{}_{\rm out} <p_1,\ldots,p_n\mid
p_1,\ldots,p_n>_{\rm in}
\,\, ,
\label{npartsmat}
\EN
factorizes into $n(n-1)/2$ two-particle $S$-matrices \cite{ZZ}
\EQ
S^{(n)}(p_1,p_2,\ldots,p_n)\,=\,\prod_{i<j} S^{(2)}_{ij}(p_i,p_j)\hs\hs\hs.
\EN
It is convenient to use instead of the momenta the rapidities $\beta_i$,
defined by
\EQ
p^0_i\,=\, m_i\cosh\,\beta_i\hs\hs,\hs\hs\hs
p^1_i\,=\, m_i\sinh\,\beta_i\hs\hs.
\label{rapidities}
\EN
By Lorentz invariance, the scattering amplitudes will be functions of the
rapidity differences $\beta_{ij}=\beta_i-\beta_j$. The two-particle $S$-matrix
satisfies the usual axioms of unitarity and crossing symmetry
\bea
S_{ij}(\be_{ij}) &=& S_{ji}(\be_{ij})  = S_{ij}^{-1}(-\be_{ij}) \,\, ,
\label{uncros} \\
S_{i\bar{\jmath}}(\be_{ij}) &=& S_{ij}(i \pi -\be_{ij}) \,\, .\nonumber
\eea
Possible bound states will occur as simple or higher odd poles in the
$S$-matrix for purely imaginary values of $\beta$ in the physical strip
$0< {\rm Im}\beta< \pi$.

Once the $S$-matrix is known, it is possible to analyze the off-shell
quantum field theory by considering the form-factors which are matrix
elements of local operators ${\cal O}(x)$ between the asymptotic states.
Pioneering work on this subject has been carried out by the authors of
ref.\,\cite{Karowski} and more recently advances have been made by
Smirnov and Kirillov \cite{nankai,Smirnov1,Smirnov2}. In order to provide
a self-consistent account of the paper, we recall some essential properties
of the form-factors, staying close to the notations of ref.\,\cite{nankai}.

\subsection{Zamolodchikov algebra}

\,

At the heart of the construction of the form factors lies the assumption that
there exists a set of vertex operators, of creation and annihilation type,
i.e. $V_{\alpha_{i}}^{\dag} (\be_i)$, $V_{\alpha_{i}}(\be_i)$, which provide
a generalization of the bosonic and fermionic algebras. Here the $\alpha_i$
denotes some quantum number indicating the different types of particles present
in the theory. These operators are assumed to obey the following non-Abelian,
associative algebra, involving the $S$-matrix
\bea
V_{\alpha_{i}}(\be_i) V_{\alpha_{j}}(\be_j) &=& S_{ij}(\be_{ij}) \,
V_{\alpha_{j}}(\be_j) V_{\alpha_{i}}(\be_i)   \label{Zamalga}     \\
V_{\alpha_{i}}^{\dag}(\be_i) V_{\alpha_{j}}^{\dag}(\be_j) &=&S_{ij}(\be_{ij})\,
V_{\alpha_{j}}^{\dag}(\be_j) V_{\alpha_{i}}^{\dag}(\be_i)  \label{Zamalgb} \\
V_{\alpha_{i}}(\be_i) V_{\alpha_{j}}^{\dag}(\be_j) &=& S_{ij}(\be_{ji}) \,
V_{\alpha_{j}}^{\dag}(\be_j) V_{\alpha_{i}}(\be_i) + 2\pi\delta_{\alpha_{i}
\alpha_{j}} \delta({\be_{ij}}) \, . \label{Zamalgc}
\eea
Each commutation of these operators is thus interpreted as a scattering
process.
The Poincar\'e group generated by the Lorentz transformation $L(\epsilon)$ and
the translation $T_y$ is expected to act on these operators in the following
way
\bea
U_L V_{\alpha}(\be)  U_L^{-1} &=&  V_{\alpha}(\be+ \epsilon)  \label{poinl}  \\
U_{T_{y}} V_{\alpha}(\be)  U_{T_{y}}^{-1} &=& e^{i p_{\mu}(\be) y^{\mu} }
V_{\alpha}(\be)     . \label{point}
\eea
Clearly, the explicit form of these vertex operators depends crucially on the
nature of the theory under consideration and a realization of such a
construction remains hitherto an open challenge for most theories.

\subsection{Physical States}

\,

We can use the vertex operators introduced in the previous section in order to
define a space of physical states. For this aim, let us consider the vacuum
$|0 \rangle$ which is the state annihilated by the operator $V_{\alpha}(\be)$,
\EQ
V_{\alpha}(\be) |0 \rangle = 0 = \langle 0 | V_{\alpha}^{\dag}(\be).
\EN
The Hilbert space is then defined by a successive action of
$V_{\alpha}^{\dag}(\be)$ on $|0 \rangle$, {\em i.e.}
\EQ
| V_{\alpha_{1}}(\be_{1}) \dots V_{\alpha_{n}}(\be_{n}) \rangle  \equiv
V_{\alpha_{1}}^{\dag}(\be_{1}) \dots V_{\alpha_{n}}^{\dag}(\be_{n}) |0 \rangle
\,\, .
\label{Hilb}
\EN
{}From equation (\ref{Zamalgc}), the one particle states are normalized as
\EQ
\langle V_{\alpha_{i}}(\be_{i}) | V_{\alpha_{j}}(\be_{j})  \rangle =\,2\pi\,
\delta_{\alpha_{i} \alpha_{j}} \delta(\be_{ij}) \, .
\EN
The algebra of the vertex operators implies that the vectors (\ref{Hilb})
are not linearly independent and in order to obtain a basis of linearly
independent states we require some additional restrictions. In \cite{ZZ} the
following prescription was proposed: Select as a basis for the in-states those
which are ordered with decreasing rapidities
\[\be_1 >\dots > \be_n
\]
and as a basis for the out-states those with increasing rapidities
\[
\be_1 < \dots < \be_n \,\,\, .
\]
These conditions select a set of linearly independent vectors which serve as
a unique basis.

\subsection{Form Factors}
\,

If not explicitly mentioned, in the following we will consider matrix elements
between in-states and out-states of hermitian local scalar operators
${\cal O}(x)$
of a theory with only one self-conjugate particle
\EQ
{}_{\rm out} \langle V(\be_{m+1}) \dots V(\be_n) |
{\cal O}(x)  | V(\be_{1}) \dots V(\be_m)
\rangle_{ \rm in} .  \label{INOUT}
\EN
Matrix elements of higher spin operators will be discussed later on. We can
always shift the matrix elements (\ref{INOUT}) to the origin by means of a
translation on the operator ${\cal O}(x)$, i.e.
$U_{T_{y}}{\cal O}(x)U_{T_{y}}^{-1} = {\cal O}( x + y)$ and by using
eq.\,(\ref{point}),
\begin{eqnarray}
&& \exp \left[i\left( \sum\limits_{i=m+1}^{n} p_{\mu} (\be_i) -
\sum\limits_{i=1}^{m} p_{\mu} (\be_i) \right) x^{\mu}\right]\, \label{matrsh}\\
&& \hspace{3mm}\times\, {}_{\rm out} \langle V(\be_{m+1}) \dots V(\be_n) |
{\cal O}(0)  | V(\be_{1}) \dots V(\be_m)
\rangle_{ \rm in} \,\,\, .\nonumber
\end{eqnarray}
It is convenient to introduce the following functions, called form factors
(fig.\,1)
\EQ
F_n^{\cal O} (\beta_1,\beta_2,\ldots,\beta_n) \,=\,
\langle 0\mid{\cal O}(0,0)\mid \beta_1,\beta_2,\ldots,\beta_n\rangle_{in} .
\label{FF}
\EN
which are the matrix elements of an operator at the origin between
an $n$-particle in-state and the vacuum\footnote{ Here and in
the following we use a more simplified notation for the physical states
$| \dots V_{\alpha'_{n}}(\be'_n) \dots \rangle \equiv | \dots \be_n
\dots \rangle$. In most cases we will also suppress the superscript
${\cal O}$ and only use it when considering form factors related
to different local operators.}. For local scalar operators ${\cal O}(x)$,
relativistic invariance implies that the form factors $F_n$ are functions
of the difference of the rapidities $\beta_{ij}$
\EQ
F_n (\beta_1,\beta_2,\ldots,\beta_n) \,=\,
F_n (\beta_{12},\beta_{13},\ldots,\beta_{ij},\ldots) \,\,\,\,, i<j
\,\,\, .
\EN
Crossing symmetry also implies that the most general matrix element
(\ref{matrsh}) is obtained by an analytic continuation of (\ref{FF}), and
equals
\EQ
F_{n+m}
(\beta_1,\beta_2,\ldots,\beta_m,\beta_{m+1}-i\pi,\ldots,\beta_n -i\pi) \,=\,
F_{n+m}(\beta_{ij},i\pi-\beta_{sr},\beta_{kl})
\label{rep}
\EN
where $1\leq i<j\leq m$, $1\leq r\leq m<s\leq n$, and $m<k<l\leq n$.

Except for the poles corresponding to the one-particle bound states in all
sub-channels, we expect the form factors $F_n$ to be analytic inside the strip
$0 < {\rm Im } \be_{ij} < 2\pi$.

\subsection{Watson's Equations}

\,

The form factors of a hermitian local scalar operator ${\cal O}(x)$ satisfy
a set of equations, known as Watson's equations \cite{Watson}, which for
integrable systems assume a particularly simple form
\bea
F_n(\be_1, \dots ,\be_i, \be_{i+1}, \dots, \be_n) &=& F_n(\be_1,\dots,\be_{i+1}
,\be_i ,\dots, \be_n) S(\beta_i-\beta_{i+1}) \,\, ,
\label{permu1}\\
F_n(\be_1+2 \pi i, \dots, \be_{n-1}, \be_n ) &=& F_n(\be_2 ,\ldots,\be_n,
\be_1) = \prod_{i=2}^{n} S(\beta_i-\beta_1) F_n(\be_1, \dots, \be_n)
\,\, .
\nonumber
\eea
The first equation is simply a consequence of (\ref{Zamalga}), i.e. as a
result of the commutation of two operators we get a scattering process.
Concerning the second equation, it states which is the discontinuity on the
cuts $\beta_{1i} = 2 \pi i$.
In the case
$n=2$, eqs.\,(\ref{permu1}) reduce to
\EQ
\begin{array}{ccl}
F_2(\beta)&=&F_2(-\beta)S_2(\beta) \,\, ,\\
F_2(i\pi-\beta)&=&F_2(i\pi+\beta) \,\,\, .
\end{array}
\label{F2}
\EN
Smirnov \cite{nankai,Smirnov2} has shown that eqs.\,(\ref{permu1}), together
with eqs.\,(\ref{recursive}) and (\ref{respole}) which will be discussed
in the next section, can be regarded as a system of axioms which defines the
whole local operator content of the theory.

The general solution of Watson's equations can always be brought into the form
\cite{Karowski}
\EQ
F_n(\beta_1,\dots,\beta_n) =K_n(\beta_1,\dots,\beta_n) \prod_{i<j}F_{\rm min}
(\beta_{ij})  \,\, ,
\label{parametrization}
\EN
where $F_{\rm min}(\beta)$ has the properties that it satisfies (\ref{F2}), is
analytic in $0\leq$ Im $\beta\leq \pi$, has no zeros in $0<$ Im $\beta<\pi$,
and converges to a constant value for large values of $\beta$. These
requirements
uniquely determine this function, up to a normalization. The remaining factors
$K_n$ then satisfy Watson's equations with $S_2=1$, which implies that they are
completely symmetric, $2 \pi i$-periodic functions of the $\beta_{i}$. They
must contain all the physical poles expected in the form factor under
consideration and must satisfy a correct asymptotic behaviour for large value
of $\beta_i$. Both requirements depend on the nature of the theory and on the
operator $\cal O$.

Postponing the discussion on the pole structure of $F_n$ to the next section,
let us notice that one condition on the asymptotic behaviour of the form
factors is dictated by relativistic invariance. In fact, a simultaneous shift
in the rapidity variables results in
\EQ
F_n^{\cal O} (\beta_1+\Lambda,\beta_2+\Lambda,\ldots,\beta_n+\Lambda) \,=\,
F_n^{\cal O} (\beta_1,\beta_2,\ldots,\beta_n) \,\, ,
\label{asymp1}
\EN
For form factors of an operator ${\cal O}(x)$ of spin $s$, the previous
equation generalizes to
\EQ
F_n^{\cal O} (\beta_1+\Lambda,\beta_2+\Lambda,\ldots,\beta_n+\Lambda) \,=\,
e^{s\Lambda}\,
F_n^{\cal O} (\beta_1,\beta_2,\ldots,\beta_n) \,\, , \label{relat}
\label{asymp2}
\EN
Secondly, in order to have a power-law bounded ultraviolet behaviour of the
two-point function of the operator ${\cal O}(x)$ (which is the case we
will consider), we have to require that the form factors behave asymptotically
at most as $\exp(k \be_i)$ in the limit $\be_i \rightarrow \infty$, with $k$
being a constant independent of $i$. This means that, once we extract from
$K_n$ the denominator which gives rise to the poles, the remaining part has to
be a symmetric function of the variables $x_i\equiv e^{\beta_i}$, with a finite
number of terms, i.e. a symmetric polynomial in the $x_i$'s. It is convenient
to introduce a basis in this functional space given by the {\em elementary
symmetric polynomials} $\sigma^{(n)}_k(x_1, \dots , x_n)$ which are generated
by \cite{Macdon}
\EQ
\prod_{i=1}^n(x+x_i)\,=\,
\sum_{k=0}^n x^{n-k} \,\sigma_k^{(n)}(x_1,x_2,\ldots,x_n).
\label{generating}
\EN
Conventionally the $\sigma_k^{(n)}$ with $k>n$ and with $n<0$ are zero.
The explicit expressions for the other cases are
\EQ
\begin{array}{l}
\sigma_0=1\hs\hs,\\
\sigma_1=x_1+x_2+\ldots +x_n\hs\hs,\\
\sigma_2=x_1x_2+x_1x_3+\ldots x_{n-1}x_n\hs\hs,\\
\vdots  \qquad \qquad \vdots  \\
\sigma_n=x_1x_2\ldots x_n\hs\hs\hs.
\end{array}
\EN
The $\sigma_k^{(n)}$ are homogeneous polynomials in $x_i$ of total degree $k$
and of degree one in each variable.

\subsection{Pole Structure and Residue Equations for the Form Factors}

\,

The pole structure of the form factors induces a set of recursive equations
for the $F_n$ which are of fundamental importance for their explicit
determination. As function of the rapidity differences $\beta_{ij}$, the form
factors $F_n$ possess two kinds of simple poles.

The first kind of singularities (which do not depend on whether or not the
model possesses bound states) arises from kinematical poles located at
$\beta_{ij}=i\pi$. They are related to the one-particle pole in a subchannel
of three-particle states which, in turn, corresponds to a crossing process of
the elastic $S$-matrix. The corresponding residues are computed by the LSZ
reduction \cite{Smirnov1,Smirnov2} and give rise to a recursive equation
between the $n$-particle and the $(n+2)$-particle form factors (fig.\hs2)
\EQ
-i\lim_{\tilde\beta \rightarrow \beta}
(\tilde\beta - \beta)
F_{n+2}(\tilde\beta+i\pi,\beta,\beta_1,\beta_2,\ldots,\beta_n)=
\left(1-\prod_{i=1}^n S(\beta-\beta_i)\right)\,
F_n(\beta_1,\ldots,\beta_n)  . \label{recursive}
\EN

The second type of poles in the $F_n$ only arise when bound states are present
in the model. These poles are located at the values of $\beta_{ij}$ in the
physical strip which correspond to the resonance angles. Let
$\beta_{ij}=i u_{ij}^k$ be one of such poles associated to the bound state
$A_k$ in the channel $A_i\times A_j$. For the $S$-matrix we have (fig.\,3)
\EQ
-i\,\lim_{\beta\rightarrow i u_{ij}^k}
(\beta-i u_{ij}^k) \,S_{ij}(\beta)\,=\, \left(\Gamma_{ij}^k\right)^2
\EN
where $\Gamma_{ij}^k$ is the three-particle vertex on mass-shell.
The corresponding residue for the $F_n$ is given by \cite{Smirnov1,Smirnov2}
\EQ
-i\lim_{\epsilon\rightarrow 0} \epsilon\,
F_{n+1}(\beta+i \overline u_{ik}^j-\epsilon,
\beta-i \overline u_{jk}^i+\epsilon,\beta_1,\ldots,\beta_{n-1})
\,=\,\Gamma_{ij}^k\,F_{n}(\beta,\beta_1,\ldots,\beta_{n-1})
\,\,\, ,
\label{respole}
\EN
where $\overline u_{ab}^c\equiv (\pi-u_{ab}^c)$. This equation establishes
a recursive structure between the $(n+1)$- and $n$-particle form factors
(fig.\,4).

\subsection{Correlation Functions from Form Factors}

\,

Once the form factors of a theory are known, the correlation functions of
local operators can be written as an infinite series over multi-particle
intermediate states. For instance, the two-point function of an operator
${\cal O}(x)$ in real Euclidean space is given by
\begin{eqnarray}
& &\langle{\cal O}(x)\,{\cal O}(0)\rangle\,=
\sum_{n=0}^{\infty}
\int \frac{d\beta_1\ldots d\beta_n}{n! (2\pi)^n}
<0|{\cal O}(x)|\beta_1,\ldots,\beta_n>_{\rm in}{}_{\rm in}
<\beta_1,\ldots,\beta_n|{\cal O}(0)|0>
\label{correlation} \nonumber \\
& &\hspace{3mm} =\,\sum_{n=0}^{\infty}
\int \frac{d\beta_1\ldots d\beta_n}{n! (2\pi)^n}
\mid F_n(\beta_1\ldots \beta_n)\mid^2 \exp \left(-mr\sum_{i=1}^n\cosh\beta_i
\right)
\end{eqnarray}
where $r$ denotes the radial distance, i.e. $r=\sqrt{x_0^2 + x_1^2}$.
All integrals are convergent and one expects a convergent series as well.
Similar expressions can be derived for multi-point correlators.

\resection{The Sinh-Gordon Theory}

\,

In this paper the model we are concerned with is the Sinh-Gordon theory,
defined by the action
\EQ
{\cal S}\,=\,\int d^2x \left[
\frac{1}{2}(\partial_{\mu}\phi)^2-\frac{m^2}{g^2}
\,\cosh\,g\phi(x)\,\right]\,\,.
\label{Lagrangian}
\EN
It is the simplest example of an affine Toda Field Theories \cite{Toda},
possessing a $Z_2$ symmetry $\phi\rightarrow -\phi$. By an analytic
continuation in $g$, i.e $g\rightarrow i g$, it can formally be mapped to
the Sine-Gordon model.

There are numerous alternative viewpoints for the Sinh-Gordon model.
First, it can be regarded either as a perturbation of the free massless
conformal action by means of the relevant operator\footnote{Although
the anomalous dimension of this operator (computed with
respect to the free conformal point), is negative, $\Delta=-g^2/8\pi$,
the resulting theory is unitary. This is due to the existence of non a
nonzero vacuum expectation values of some of the fields ${\cal O}_i$ in
the theory. A detailed discussion of this point can be found in \cite{YLZam}.}
$\cosh\,g\phi(x)$. Alternatively, it can be considered as a perturbation of
the conformal Liouville action
\EQ
{\cal S}\,=\,\int d^2x \left[
\frac{1}{2}(\partial_{\mu}\phi)^2-\lambda e^{g\phi} \right]
\label{Liouville}
\EN
by means of the relevant operator $e^{-g\phi}$ or as a conformal affine
$A_1$-Toda Theory \cite{Babon} in which the conformal symmetry is broken by
setting the free field to zero.

Furthermore, it is interesting to notice that the Sinh-Gordon model can be
mapped into a Coulomb Gas system with an integer set of charges. To illustrate
this, let us consider the (Euclidean) partition function of the model
\EQ
Z(m,g)\,=\,\int {\cal D}\phi \, e^{-{\cal S}}\,\,\ .
\label{path}
\EN
Using the identity
\EQ
\exp\left(-\frac{m^2}{g^2}\cosh\,g\phi(x)\right)\,=\,
\sum_{n(x)=-\infty}^{+\infty} I_{n(x)}\left(-\frac{m^2}{g^2}\right)
\exp\left(g\,n(x)\phi(x)\right)\,\,\,,
\EN
where $I_n(a)$ denotes the Bessel function of integer order $n$, the functional
integral in (\ref{path}) becomes Gaussian and can be performed explicitly.
Hence $Z(m,g)$ can be cast in the following form
\EQ
Z(m,g)\,=\,Z(0)\,
\sum_{n(x)=-\infty}^{+\infty} I_{n(x)}\left(-\frac{m^2}{g^2}\right)\, \,
\exp\left[\frac{g^2}{2} \int dx dy \, n(x) \Delta(x-y)\,n(y)\right] \,\,\,,
\EN
where $Z(0)$ is the partition function of a massless free theory and
$\Delta(x-y)$ is the two-dimensional massless propagator. The model
is therefore equivalent to a Coulomb Gas system with integer charges and
with weight functions for the configurations given by the $I_n$'s.

In a perturbative approach to the quantum field theory defined by the
action (\ref{Lagrangian}), the only ultraviolet divergences which
occur in any order in $g$ come from tadpole graphs and can be removed
by a normal ordering prescription with respect to an arbitrary mass scale
$M$. All other Feynman graphs are convergent and give rise to
finite wave function and mass renormalisation. The coupling constant $g$
does not renormalise.

An essential feature of the Sinh-Gordon theory is its integrability, which
in the classical case can be established by means of the inverse scattering
method \cite{Faddev}. In order to obtain the expressions of the (classical)
conserved currents, let us consider the Euclidean version of the model in
terms of the complex coordinates $z$ and $\overline z$
\EQ
z\,=\,(x^0+i x^1)\,\,;
\hspace{3mm}
\overline z\,=\,(x^0-i x^1)\,\, ,
\EN
and define a field $\hat \phi(z,\overline z,\epsilon)$ which satisfies
the following (B\"acklund) equations
\begin{eqnarray}
\frac{\partial}{\partial z}(\hat\phi+\phi)\,=\,
\frac{m}{g}\epsilon \,\sinh\left(\frac{g}{2}(\hat\phi-\phi)\right)\,\, ,
\label{Backlund}\\
\frac{\partial}{\partial \overline z}(\hat\phi-\phi)\,=\,
\frac{m}{g\epsilon} \,\sinh\left(\frac{g}{2}(\hat\phi+\phi)\right)
\,\, . \nonumber
\end{eqnarray}
Given that $\phi(z,\overline z)$ is a solution of the equation of motion
originated by (\ref{Lagrangian}), eqs.\,(\ref{Backlund}) define a new solution
$\hat\phi(z,\overline z,\epsilon)$ and imply as well the following
conservation laws
\EQ
\epsilon^{-1}\,\partial_{z}\left(\cosh\,\frac{g}{2}(\hat\phi+\phi)\right)
-\epsilon\,\partial_{\overline z}
\left(\cosh\,\frac{g}{2}(\hat\phi-\phi)\right)=0\hs\hs\hs.
\label{conservation}
\EN
$\hat\phi(z,\overline z,\epsilon)$ can be expressed in terms of a power
series in $\epsilon$
\EQ
\hat\phi(z,\overline z,\epsilon)\,=\,\sum_{n=0}^{\infty}
\phi^{(n)}(z,\overline z)\,\epsilon^n
\label{series}
\EN
with the fields $\phi^{(n)}(z,\overline z)$ calculated by using
eqs.\,(\ref{Backlund}). Placing (\ref{series}) into (\ref{conservation})
and matching equal power in $\epsilon$, one obtains an infinite set of
conservation laws
\EQ
\partial_{z}\,T_{s+1}\,=\,
\partial_{\overline z}\,\Theta_{s-1}\,\, .
\EN
The corresponding charges ${\cal Q}_s$ are given by
\EQ
{\cal Q}_s\,=\,\oint \left[ T_{s+1}\,dz+\Theta_{s-1}\,d\overline z\right]
\,\,\, .
\EN
The integer-valued index $s$ which labels the integrals of motion is the spin
of the operators. Non-trivial conservation laws are obtained for odd values of
$s$
\EQ
s=1,3,5,7,\ldots
\label{conservedspin}
\EN
In analogy to the Sine-Gordon theory \cite{Sazaki}, an infinite set of
conserved charges ${\cal Q}_s$ with spin $s$ given in (\ref{conservedspin})
also exists for the quantized version of the Sinh-Gordon theory. They are
diagonalised by the asymptotic states with eigenvalues given by
\EQ
{\cal Q}_s \,|\beta_1,\ldots,\beta_n>\,=\,
\chi_s\,\sum_{i=1}^n e^{s\beta_i}\, |\beta_1,\ldots,\beta_n>\,\,\, ,
\label{charges}
\EN
where $\chi_s$ is the normalization constant of the charge ${\cal Q}_s$. The
existence of these higher integrals of motion precludes the possibility of
production processes and hence guarantees that the $n$-particle scattering
amplitudes are purely elastic and factorized into $n(n-1)/2$ two-particle
$S$-matrices. The exact expression for the Sinh-Gordon theory is given by
\cite{AFZ}
\EQ
S(\beta,B)\,=\,
\frac{\tanh\frac{1}{2}(\beta-i\frac{\pi B}{2})}
{\tanh\frac{1}{2}(\beta+i\frac{\pi B}{2})} \label{smatrix}\,\, ,
\EN
where $B$ is the following function of the coupling constant $g$
\EQ
B(g)\,=\,\frac{2g^2}{8\pi+g^2} \hs\hs\hs.
\EN
This formula has been checked against perturbation theory in ref.\,\cite{AFZ}
(more recently to higher orders in \cite{Braden}) and can also be obtained by
analytic continuation of the $S$-matrix of the first breather of the
Sine-Gordon theory \cite{ZZ}. For real values of $g$ the $S$-matrix has no
poles in the physical sheet and hence there are no bound states, whereas
two zeros are present at the crossing symmetric positions
\EQ
\beta\,=\,
\left\{
\begin{array}{l}
i\frac{\pi B}{2} \\
i\frac{\pi (2-B)}{2}
\end{array}
\right.
\EN
The absence of bound states in the Sinh-Gordon model is also supported by the
general fusing rule of affine Toda Field Theories \cite{fuse}.

An interesting feature of the $S$-matrix is its invariance under the map
\cite{MC}
\EQ
B\rightarrow 2-B \label{mapd}
\EN
{\em i.e.} under the {\em strong-weak} coupling constant duality
\EQ
g\rightarrow \frac{8\pi}{g} .\label{mapdu}
\EN
This duality is a property shared by the unperturbed conformal Liouville theory
(\ref{Liouville}) \cite{Mansfield} and it is quite remarkable that it survives
even when the conformal symmetry is broken.

\section{Form Factors for the Sinh-Gordon Theory}

\,

The $Z_2$ symmetry of the model is realized by a map $\sigma$, whose effect
on the elementary field of the theory is $\sigma(\phi) = -\phi$. We assume
that it has the same effect on the vertex operator, that is
$\sigma(V(\beta)) = -V(\beta)$ together with $\sigma( V(\beta_1)
V(\beta_2) )
= \sigma( V(\beta_1) ) \sigma( V(\beta_2))$. According to this symmetry we can
label the operators by their $Z_2$ parity.

For operators which are $Z_2$-odd the only possible non-zero form factors are
those involving an odd number of particles, i.e.
\EQ
F^{\cal O}_{2n}(\be_1, \dots , \be_{2n}) = 0 \qquad \hbox{for} \qquad
\sigma({\cal O}) = - {\cal O} .
\EN
This implies in particular that ${\cal O}$ cannot acquire a non-zero vacuum
expectation value. On the other hand, for $Z_2$-even operators the only
possible non-zero form factors are those involving an even number of
particles, i.e.
\EQ
F^{\cal O}_{2n+1}(\be_1, \dots , \be_{2n+1}) = 0 \qquad \hbox{for} \qquad
\sigma({\cal O}) =  {\cal O} .
\EN
The vacuum expectation value of $Z_2$-even operators can in principle
be different from zero.

The simplest representative of the odd sector is given by the (renormalised)
field $\phi(x)$ itself. It creates a one-particle state from the vacuum.
Our normalization is fixed to be (see sect.\,4.3)
\EQ
F_1^{\phi}(\beta)\,=\, <0\mid \phi(0) \mid \beta>_{\rm in}\,=\,
\frac{1}{\sqrt{2}} \,\, .
\label{normphi}
\EN
For the even sector, an important operator is given by the energy-momentum
tensor
\EQ
T_{\mu\nu}(x)\,=\,2\pi\,(:\partial_{\mu}\phi \partial_{\nu}\phi
-g_{\mu\nu} {\cal L}(x)\,:)
\EN
where $:\,:$ denotes the usual normal ordering prescription with respect
an arbitrary mass scale $M$. Its trace $T_{\mu}^{\,\mu}(x)=\Theta(x)$ is a
spinless operator whose normalization is fixed in terms of its two-particle
form factor
\EQ
F_2^{\Theta}(\beta_{12}=i\pi)\,=\, {}_{\rm out}<\beta_1\mid
\Theta(0) \mid \beta_2>_{\rm in}\,=\,2 \pi m^2
\,\, ,
\label{normtheta}
\EN
where $m$ is the physical mass.

In the following we shall compute the form factors of the operators $\phi(x)$
and $\Theta(x)$. This will be sufficient to characterize the basic properties
of the model since form factors for other operators can in general be obtained
from simple arguments once $F_{n}^{\phi}$ and $F_{n}^{\Theta}$ are known. For
instance, suppose we want to compute the form factors of the operator
${\cal O} = :\sinh\, g \phi:$. They can be easily computed in terms of the
form factor for $\phi$. In fact, using eq.\,(\ref{matrsh}) we have
\EQ
\langle 0 | \partial_z \partial_{\zb} \phi(z,\zb) | \be_1 \dots
\be_m\rangle _{\rm in} = - {{m^2} \over 4} \sum_i e^{\be_i}
\sum_i e^{-\be_i}  \sum_i  e^{-i x p_i } F_n^{\phi}( \be_1, \dots,\be_n)
\EN
Employing the equation of motion and choosing $ z=\zb=0$, together with
the identities
\EQ
\sum_{i=1}^n e^{\be_{i}} = \sigma^{(n)}_1 (x_1, \dots, x_n)  \qquad
, \qquad \sum_{i=1}^n e^{-\be_{i}} = {{\sigma^{(n)}_{n-1}(x_1, \dots,
x_n) } \over {\sigma^{(n)}_{n} (x_1, \dots, x_n) }}
\label{identita}
\EN
we derive the relation
\EQ
\sigma_n F_n^{\sinh\,g \phi} = g \sigma_1 \sigma_{n-1} F_n^{\phi}\,\,\, .
\EN
In the last section of this paper we also describe how form factors of
operators with higher spin arise from the knowledge of $F_n^{\Theta}$ or
$F_n^{\phi}$.

\subsection{Minimal Two-Particle Form Factor}

\,

An essential step for the computation the form factors is the determination of
$F_{\rm min}(\beta)$, introduced in (\ref{parametrization}). It satisfies
the equations
\EQ
\begin{array}{ccl}
F_{\rm min}(\beta)&=&F_{\rm min}(-\beta)\, S_2(\beta)\,\, ,\\
F_{\rm min}(i\pi-\beta)&=&F_{\rm min}(i\pi+\beta)\,\,\, .
\end{array}
\label{Watson2}
\EN
As shown in \cite{Karowski}, the easiest way to compute $F_{\rm min}(\beta)$
(up to a normalization ${\cal N}$) is to exploit an integral representation
of the $S$-matrix
\EQ
S(\beta)\,=\, \exp\,\left[\int_0^{\infty}\frac{dx}{x} f(x)
\sinh\left(\frac{x\beta}{i\pi}\right)\right]
\,\,\, .
\EN
Then a solution of (\ref{Watson2}) is given by
\EQ
F_{\rm min}(\beta)\,=\,{\cal N}\,\exp\,
\left[\int_0^{\infty} \frac{dx}{x} f(x)
\frac{\sin^2\left(\frac{x\hat\beta}{2\pi}\right)}{\sinh x}\right] \,\,\,.
\EN
where
\EQ
\hat\beta\equiv i\pi-\beta\,\,\, .
\EN
For the Sinh-Gordon theory we have
\EQ
F_{\rm min}(\beta,B)\,=\,{\cal N}\,\exp\,\left[
8\int_0^{\infty} \frac{dx}{x} \frac{\sinh\left(\frac{x B}{4}\right)
\sinh\left(\frac{x}{2}(1-\frac{B}{2})\right) \,\sinh\frac{x}{2}}{\sinh^2 x}
\sin^2\left(\frac{x\hat\beta}{2\pi}\right)\right] \,\,\,.
\label{integral}
\EN
We choose our normalization to be
\EQ
{\cal N} \,=\,\exp\left[-4\int_0^{\infty}
\frac{dx}{x} \frac{\sinh\left(\frac{x B}{4}\right)
\sinh\left(\frac{x}{2}(1-\frac{B}{2})\right) \,\sinh\frac{x}{2}}{\sinh^2 x}
\right] \,\,\,.
\EN
The analytic structure of $F_{\rm min}(\beta,B)$ can be easily read from
its infinite product representation in terms of $\Gamma$ functions
\EQ
F_{\rm min}(\beta,B)\,=\,
\prod_{k=0}^{\infty}
\left|
\frac{\Gamma\left(k+\frac{3}{2}+\frac{i\hat\beta}{2\pi}\right)
\Gamma\left(k+\frac{1}{2}+\frac{B}{4}+\frac{i\hat\beta}{2\pi}\right)
\Gamma\left(k+1-\frac{B}{4}+\frac{i\hat\beta}{2\pi}\right)}
{\Gamma\left(k+\frac{1}{2}+\frac{i\hat\beta}{2\pi}\right)
\Gamma\left(k+\frac{3}{2}-\frac{B}{4}+\frac{i\hat\beta}{2\pi}\right)
\Gamma\left(k+1+\frac{B}{4}+\frac{i\hat\beta}{2\pi}\right)}
\right|^2
\EN
$F_{\rm min}(\beta,B)$ has a simple zero at the threshold $\beta=0$ since
$S(0)=-1$ and its asymptotic behaviour is given by
\EQ
\lim_{\beta \rightarrow \infty} F_{\rm min}(\beta,B) = 1.
\EN
It satisfies the functional equation
\EQ
F_{\rm min}(i\pi+\beta,B) F_{\rm min}(\beta,B)\,=\,
\frac{\sinh\beta}{\sinh\beta+\sinh\frac{i\pi B}{2}}
\label{shift}
\EN
which we will use in the next section in order to find a convenient form
for the recursive equations of the form factors.

A useful expression for the numerical evaluation of $F_{\rm min}(\beta,B)$
is given by
\begin{eqnarray}
& &F_{\rm min}(\beta,B) \,=\, {\cal N}
\prod_{k=0}^{N-1}
\left[\frac{\left(1+\left(\frac{\hat\beta/2 \pi}{k+\frac{1}{2}}\right)^2\right)
\left(1+\left(\frac{\hat\beta/2 \pi}{k+\frac{3}{2}-\frac{B}{4}}\right)^2\right)
\left(1+\left(\frac{\hat\beta/2 \pi}{k+1+\frac{B}{4}}\right)^2\right)}
{\left(1+\left(\frac{\hat\beta/2 \pi}{k+\frac{3}{2}}\right)^2\right)
\left(1+\left(\frac{\hat\beta/2 \pi}{k+\frac{1}{2}+\frac{B}{4}}\right)^2\right)
\left(1+\left(\frac{\hat\beta/2 \pi}{k+1-\frac{B}{4}}\right)^2\right)}\right]^
{k+1} \\
& &\times \,
\exp\,\left[
8\int_0^{\infty} \frac{dx}{x} \frac{\sinh\left(\frac{x B}{4}\right)
\sinh\left(\frac{x}{2}(1-\frac{B}{2})\right) \,\sinh\frac{x}{2}}{\sinh^2 x}
(N+1-N\,e^{-2x})\,e^{-2Nx}\,
\sin^2\left(\frac{x\hat\beta}{2\pi}\right)\right] \,\,\,.\nonumber
\end{eqnarray}
The rate of convergence of the integral may be improved substantially
by increasing the value of $N$. Graphs of $F_{\rm min}(\beta,B)$ are
drawn in fig.\,5.

\vspace{1cm}

\subsection{Parametrization of the $n$-Particle Form Factors}

\,

Since the Sinh-Gordon theory has no bound states, the only poles
which appear in any form factor $F_n(\beta_1,\ldots,\beta_n)$ are those
occurring in every three-body channel. Additional poles in the $n$-body
intermediate channel are excluded by the elasticity of the scattering theory.
Using the identity
\EQ
(p_1+p_2+p_3)^2-m^2\,=\,8m^2 \cosh\frac{1}{2}\beta_{12}
\cosh\frac{1}{2}\beta_{13} \cosh\frac{1}{2}\beta_{23}\,\, ,
\EN
all possible three-particle poles are taken into account by the following
parameterization of the function
\EQ
K_n(\beta_1,\dots,\beta_n) \,=\,\frac{Q'_n(\beta_1,\dots,\beta_n)}
{\prod\limits_{i<j}\cosh\frac{1}{2}\beta_{ij}}\,\, ,
\EN
where $Q'_n$ is free of any singularity. The second equation in (\ref{permu1})
implies that $Q'_n$ is $2 \pi i$-periodic (anti-periodic) when $n$ is an odd
(even) integer. Hence, with a re-definition of $Q'_n$ into $Q_n$, the general
parameterization of the form factor $F_n(\beta_1,\ldots \beta_n)$ is chosen to
be
\EQ
F_n(\beta_1,\ldots,\beta_n)\,=\, H_n\, Q_n(x_1,\ldots,x_n)\,
\prod_{i<j} \frac{F_{\rm min}(\beta_{ij})}{x_i+x_j}
\,\label{para}
\EN
where $x_i=e^{\beta_i}$ and $H_n$ is a normalization constant. The denominator
in (\ref{para}) may be written more concisely as $\det \Sigma$ where the
entries of the $(n-1) \times (n-1)$-matrix $\Sigma$ are given by
$\Sigma_{ij}=\sigma^{(n)}_{2i-j}(x_1,\dots, x_n)$.

The functions $Q_n(x_1,\dots,x_n)$ are symmetric polynomials in the variables
$x_i$. As consequence of eq.\,(\ref{asymp1}), for form factors of spinless
operators the total degree should be $n(n-1)/2$ in order to match the total
degree of the denominator in (\ref{para}). Form factors of higher spin
operators will be considered in sect.\,6. The order of the degree of $Q_n$ in
each variable $x_i$ is fixed by the nature and by the asymptotic behaviour
of the operator $\cal O$ which is considered.

Employing now the parameterization (\ref{para}), together with the identity
(\ref{shift}), the recursive equations (\ref{recursive}) take on the form
\EQ
(-)^n\,Q_{n+2}(-x,x,x_1,\ldots,x_n)\, = \,x  D_n(x,x_1,x_2,\ldots ,x_n)
\,Q_n(x_1,x_2,\ldots,x_n)
\label{gleich}
\EN
where we have introduced the function
\EQ
D_n=\, {{-i} \over {4\sin(\pi B/2)}} \left(\prod_{i=1}^n
\left[(x+\omega x_i)(x-\omega^{-1}x_i)\right] - \prod_{i=1}^n \left[
(x-\omega x_i)(x+\omega^{-1}x_i)\right]\right)
\EN
with $\omega=\exp(i\pi B/2)$. The normalization constants for the
form factors of odd and even operators are conveniently chosen to be
\bea
H_{2n+1} &=& H_1 \left(\frac{4\sin(\pi B/2)}{ F_{\rm min}(i \pi,B)}
\right)^n \\
H_{2n} &=& H_2 \left(\frac{4\sin(\pi B/2)}{ F_{\rm min}(i \pi,B)}
\right)^{n-1}\nonumber
\eea
where $H_1$ and $H_2$ are the initial conditions, fixed by the nature
of the operator. Using the generating function (\ref{generating}) of the
symmetric polynomials, the function $D_n$ can be expressed as
\EQ
D_n={{1} \over {2 \sin(\pi B/2)}} \sum_{l,k=0}^n (-1)^l \sin\left(
(k-l) {{\pi B} \over{2}} \right) x^{2n -l-k} \sigma_l^{(n)} \sigma_k^{(n)}.
\label{sum}
\EN
As function of $B$, $D_n$ is invariant under $B\rightarrow -B$. The non-zero
terms entering the sum (\ref{sum}) are those involving the ratios
\[
\frac{\sin(n\,\pi B/2)}{\sin(\pi B/2)}
\]
$n$ being an odd number. This means that $D_n$ may only contain powers
of $\cos^2(\pi B/2)$.

\subsection{LSZ Formula for Form Factors}

\,

The aim of this section is to show that the symmetric polynomials $Q_{2n+1}$
entering the form factors of the elementary field $\phi(x)$ can be factorized
as
\EQ
Q_{2n+1}(x_1,\ldots,x_{2n+1})\,=\,\sigma_{2n+1}^{(2n+1)} \,
P_{2n+1}(x_1,\ldots,x_{2n+1}) \hspace{5mm} n>0 \,\,\, ,
\label{field}
\EN
whereas the analogous polynomials entering the form factors of the trace of the
stress-energy tensor can be written as
\EQ
Q_{2n}(x_1,\ldots,x_{2n})\,=\,\sigma_1^{(2n)} \sigma_{2n-1}^{(2n)} \,
P_{2n}(x_1,\ldots,x_{2n}) \hspace{5mm} n>1 \,\,\, .
\label{trace}
\EN
$P_n(x_1,\ldots,x_n)$ is a symmetric polynomial of total degree $n(n-3)/2$
and of degree $n-3$ in each variable $x_i$. Using the following property of
the elementary symmetric polynomials
\EQ
\sigma_k^{(n+2)}(-x,x,x_1,\ldots,x_n)\,=\,
\sigma_k^{(n)}(x_1,x_2,\ldots,x_n)-x^2 \sigma_{k-2}^
{(n)}(x_1,x_2,\ldots,x_n)  \,\,\, ,
\label{kinshift}
\EN
the recursive equations (\ref{gleich}) can then be written in terms of the
$P_n$ as
\EQ
(-)^{n+1}\,P_{n+2}(-x,x,x_1,\ldots,x_n)\, = \,
\frac{1}{x} D_n(x,x_1,x_2,\ldots ,x_n)
\,P_n(x_1,x_2,\ldots,x_n) \,\, .
\label{gleich1}
\EN

In order to show the factorization (\ref{field}) for $F_{2n+1}^{\phi}$, it is
useful to recall the LSZ formula for the form factors of a local operator
${\cal O}(x)$
\EQ
F_n(\beta_1,\beta_2,\ldots,\beta_n)\,=\,
\left({{1} \over {\sqrt{2}}}\right)^n\,\lim_{p_{i}^2\,\rightarrow m^2}\,
\prod_{i=1}^n \left({p_{i}^2 - m^2 \over {i}}\right)\,
G^{n,{\cal O}} (q= - \sum_{i=1}^n p_{i}, p_{1}, p_{2}, \ldots, p_{n})
\label{LSZ}
\EN
where
\begin{eqnarray}
 \lefteqn{
(2 \pi)^2 \delta^2 (q + \sum p_{i})
G^{n,{\cal O}} (q, p_{1}, p_{2}, \ldots, p_{n})\,=} \nonumber \\
 & &
\int \prod_{i=1}^n dx_i \,dy e^{- i \sum p_i x_i} e^{- i q y}
<0\mid T({\cal O}(y)\phi(x_1) \phi(x_2) \ldots \phi(x_n) )\mid0>\,\,\,.
\label{FTGF}
\end{eqnarray}
The utility of these equations is threefold. First they may allow us to fix the
initial condition of the recursive equations (\ref{recursive}). Second,
they permit to study the asymptotic behaviour of the form factors, with
a corresponding restriction on the space of solutions. Finally, they provide
a tool to check our result through perturbation theory.

When ${\cal O}(x)$ is the field $\phi(x)$ itself, the application of
eq.\,(\ref{LSZ}) for $n=1$ gives
\EQ
F_1^{\phi}(\beta)\,=\, <0\mid \phi(0) \mid \beta>_{\rm in}\,=\,
\frac{1}{\sqrt{2}} \lim_{p^2 \rightarrow m^2} \frac{p^2 - m^2}{i}
G^2(p)
\EN
which provides the initial condition
\EQ
F_1^{\phi}(\beta)\,=\,\frac{1}{\sqrt{2}}\,\, .
\EN
It is now easy to establish that the form factors $F_{2n+1}^{\phi}$ of the
elementary field $\phi(x)$ are proportional to $\sigma_{2n+1}^{(2n+1)}$. The
reason is that from any Feynman diagram which enters $F_{2n+1}^{\phi}$ we can
factorize the propagator
\EQ
\frac{i}{q^2-m^2}\left|_{q=-\sum p_i {\rm ,}\,\,p_i^2=m^2}
\right.
\label{propagator}
\EN
that, written in terms of the variables $x_i$, becomes proportional to
$\sigma_{2n+1}^{(2n+1)}$
\EQ
\frac{i}{q^2-m^2}\mid_{q=-\sum p_i {\rm ,}\,\,p_i^2=m^2}\,=\,
\frac{i}{m^2} \frac{\sigma_{2n+1}^{(2n+1)}}{\sigma_1^{(2n+1)}
\sigma_{2n}^{(2n+1)} - \sigma_{2n+1}^{(2n+1)}}\,\, .
\EN
The presence of the propagator (\ref{propagator}) in front of any form factor
of the elementary field $\phi(x)$ also implies that $F_{2n+1}^{\phi}$ behaves
asymptotically as
\EQ
F_{2n+1}(\beta_1,\beta_2,\ldots,\beta_{2n+1})\rightarrow 0 \,\,\,\,{\rm as}\,\,
\beta_i \rightarrow + \infty \,\,\,\,\beta_{j \neq i} \,\,{\rm fixed}.
\label{ASYMP}
\EN
In fact, the propagator (\ref{propagator}) goes to zero in this limit whereas
the remaining expression of the Feynman graphs entering $F_{2n+1}$ is a
perturbative series which starts from the tree level vertex diagram shown in
fig.\,6, which is a constant. Other tree level contributions at the lowest
order and higher order corrections are either finite or they vanish in the
limit (\ref{ASYMP}). In fact, by dimensional analysis they must have external
momenta in the denominator in order to compensate the increasing power of the
mass in the coupling constants.

In order to prove the factorization (\ref{trace}) for the form factors
$F_{2n}^{\Theta}$ of the trace of the stress-energy tensor, let us consider
the conservation laws satisfied by this operator
\EQ
\partial_{\zb} T(z,\zb) + \partial_z \Theta(z,\zb) = 0    \hspace{3mm},
\qquad \qquad
\partial_z \bar{T}(z,\zb) + \partial_{\zb} \Theta(z,\zb) = 0
\label{conser}
\EN
where $T$ ($\bar{T}$) is the component of the stress-energy tensor which in
the conformal limit becomes holomorphic (anti-holomorphic). Using
eq.\,(\ref{matrsh}), the identities (\ref{identita} together with
(\ref{conser}), we obtain
\bea
\sigma^{(2n)}_1 \sigma^{(2n)}_{2n} F_{2n}^T (\be_1, \dots, \be_{2n} ) &=&
\sigma^{(2n)}_{2n-1} F_{2n}^{\Theta} (\be_1, \dots, \be_{2n} ) \\
\sigma^{(2n)}_{2n-1} F_{2n}^{\bar{T}} (\be_1, \dots, \be_{2n} ) &=&
\sigma^{(2n)}_1 \sigma^{(2n)}_{2n} F_{2n}^{\Theta} (\be_1, \dots, \be_{2n} ) .
\eea
Since $F_{2n}^T$, $F_{2n}^{\bar{T}} $ and $F_{2n}^{\Theta}$ are expected
to have the same analytic structure, we conclude that $F_{2n}^{\Theta}
(\be_1, \dots, \be_{2n} )$ is proportional to the product $\sigma^{(2n)}_1
\sigma^{(2n)}_{2n-1}$  for $n >2$.

\subsection{Solutions of the Recursive Equations}

\,

Let us summarize the analysis carried out in the previous sections. The form
factors $F_{2n+1}^{\phi}$ ($n>0$) of the elementary field $\phi(x)$ are given
by
\EQ
F_{2n+1}^{\phi}(\beta_1,\ldots,\beta_{2n+1})\,=\,
\frac{1}{\sqrt 2} \left(\frac{4\sin(\pi B/2)}{F_{\rm min}(i\pi,B)}\right)^n\,
\sigma_{2n+1}^{(2n+1)} P_{2n+1}(x_1,\ldots,x_{2n+1}) \,
\prod_{i<j} \frac{F_{\rm min}(\beta_{ij})}{x_i+x_j}
\EN
and the normalization of the field is fixed by
\EQ
F_{1}^{\phi}\,=\,\frac{1}{\sqrt 2} \,\,\, .
\EN
The form factors $F_{2n}^{\Theta}$ ($n>1$) of the trace of the stress-energy
tensor $\Theta(x)$ are given by
\EQ
F_{2n}^{\Theta}(\beta_1,\ldots,\beta_{2n})\,=\,
\frac{2\pi m^2}{F_{\rm min}(i\pi)}\,
\left(\frac{4\sin(\pi B/2)}{F_{\rm min}(i\pi)}\right)^{n-1}\,
\sigma_1^{(2n)} \sigma_{2n-1}^{(2n)} P_{2n}(x_1,\ldots,x_{2n}) \,
\prod_{i<j} \frac{F_{\rm min}(\beta_{ij})}{x_i+x_j}
\label{FFTHETA}
\EN
where the normalization is fixed by the matrix element of $\Theta(0)$ between
the two-particle state and the vacuum
\EQ
F_2^{\Theta}(\beta_{12})\,=\,
2 \pi m^2 \frac{F_{\rm min}(\beta_{12})}{F_{\rm min}(i \pi)} \,\,\, .
\EN
Notice that (\ref{FFTHETA}) for $n=0$ leads to the expectation value of
$\Theta$ on the vacuum
\EQ
<0|\Theta(0)|0>\,=\,\frac{\pi m^2}{2\sin(\pi B/2)} \,\,\, .
\EN
Using the recursive equations (\ref{gleich1}) and the transformation
property of the elementary symmetric polynomials (\ref{kinshift}), the
explicit expressions of the first polynomials $P_n(x_1,\ldots,x_n)$ are
given by\footnote{The upper index of the elementary symmetric
polynomials entering $P_n$ is equal to $n$ and we suppress it, in order to
simplify the notation.}
\bea
P_3(x_1,\ldots,x_3)         &=& 1 \nonumber  \\
P_4(x_1,\ldots,x_4) &=& \sigma_2\nonumber \\
P_5(x_1,\ldots,x_5) &=&  \sigma_2 \sigma_3 -c_1^2
                              \sigma_5 \\
P_6(x_1,\ldots,x_6) &=& \sigma_3 (\sigma_2\sigma_4 -\sigma_6)  -c_1^2
(\sigma_4 \sigma_5 + \sigma_1 \sigma_2 \sigma_6)  \nonumber \\
P_7(x_1,\ldots,x_7) &=& \sigma_2 \sigma_3 \sigma_4  \sigma_5  -c_1^2
(\sigma_4 \sigma_5^2 +\sigma_1 \sigma_2\sigma_5\sigma_6+
\sigma_2^2\sigma_3-c_1^2 \sigma_2\sigma_5)+ \nonumber\\
&& -c_2 (\sigma_1\sigma_6\sigma_7+\sigma_1\sigma_2\sigma_4\sigma_7+
\sigma_3\sigma_5\sigma_6)+c_1 c_2^2 \sigma_7^2 \nonumber
\eea
where $c_1=2 \cos (\pi B/2)$ and  $c_2 = 1 - c_1^2$. Expression of the higher
$P_n$ are easily computed by an iterative use of eqs.\,(\ref{gleich}).
For practical application the first representatives of $P_n$ are sufficient to
compute with a high degree of accuracy the correlation functions of the fields.
In fact, the $n$-particle term appearing in the correlation function of the
fields (\ref{correlation}) behaves as $e^{-n(mr)}$ and for quite large
values of $mr$ the correlator is dominated by the lowest number of particle
terms. This conclusion is also confirmed by an application of the $c$-theorem
which is discussed in sect.\,6. Nevertheless, it is interesting to notice that
closed expressions for $P_n$ can be found for particular values of the
coupling constant, as we demonstrate in the next subsections.

\subsubsection{The Self-Dual Point}

\,

The self-dual point in the coupling constant manifold has the special value
\EQ
B\left( \sqrt{8 \pi}\right) \, =\, 1 \,\, .
\EN
The two zeros of the $S$-matrix merge together and the function
$D_n(x,x_1,x_2,\ldots,x_n)$ acquires the particularly simple form
\EQ
D_n(x|x_1,x_2\ldots,x_n)\,=\,
\left(\sum_{k=0}^n (-1)^{k+1} \sin\frac{k\pi}{2} x^{n-k} \sigma_k^{(n)}
\right)\,
\left(\sum_{l=0}^n (-1)^{l} \cos\frac{l\pi}{2} x^{n-l} \sigma_l^{(n)}
\right)\,\hs\hs\hs. \label{recfun}
\EN
In this case the general solution of the recursive equations
(\ref{gleich1}) is given by
\EQ
P_n(x_1,x_2,\dots,x_n) \,=\,
{\rm det}\, {\cal A}(x_1,x_2,\ldots,x_n) \label{loesung}
\EN
where ${\cal A}$ is an $(n-3)\times (n-3)$ matrix whose entries are
\EQ
{\cal A}_{ij}(x_1,x_2,\ldots,x_n)\,=\,
\sigma^{(n)}_{2j-i+1}\,\cos^2\left[(i-j)\frac{\pi}{2}
\right] \,\,\, ,
\EN
i.e.
\EQ
{\cal A} = \left( \begin{array}{lllll}
\sigma_2 &    0     & \sigma_6 &   0      & \cdots \\
   0     & \sigma_3 &    0     & \sigma_7 & \cdots \\
   1     &    0     & \sigma_4 &   0      & \cdots \\
   0     & \sigma_1 &    0     & \sigma_5 & \cdots \\
 \vdots  &  \vdots  &  \vdots  &  \vdots  & \ddots \\
\end{array} \right)
\EN
This can be proved by exploiting the properties of determinants. i.e.
their invariance under linear combinations of the rows and the columns. Let us
consider the $(n-1)\times (n-1)$ matrix associated to
$P_{n+2}(-x,x,x_1,\ldots,x_n)$
\EQ
{\cal A}_{ij}\,=\,\left(\sigma^{(n)}_{2j-i+1}-x^2\,\sigma^{(n)}_{2j-i-1}\right)
\,\cos^2\left[(i-j)\frac{\pi}{2}
\right] \,\, ,
\label{initial}
\EN
where eq.\,(\ref{kinshift}) was used. Adding successively $x^2$ times the row
$(i+2)$ to row $i$ (starting with $i=1$), we obtain for the entries of the
matrix ${\cal A}$
\EQ
{\cal A}_{ij}\,=\,\left(\sigma^{(n)}_{2j-i+1}-x^4\,
\sigma^{(n)}_{2j-i-3}\right) \,\cos^2\left[(i-j)\frac{\pi}{2}
\right].
\EN
Adding now $x^4$ times of the $i^{th}$ column to column $(i+2)$ (starting
with $i=1$), we obtain the following matrix:
\EQ
{\cal A}^{(n-1) \times (n-1)} =
\left(
\begin{array}{ccc}
                             & 0                     &          0     \\
{\cal A}^{(n-3)\times (n-3)} & \vdots                &   \vdots       \\
                             &  0                    &   \vdots       \\
   \ast \cdots \ast          & {\cal A}_{(n-2)(n-2)} &  0             \\
   \ast \cdots \ast          &  0                &{\cal A}_{(n-1)(n-1)} \\
\end{array}
\right)
\EN
where the entries in the lower right corner are given by
\begin{eqnarray}
{\cal A}_{(n-2)(n-2)} & = & \sum_{k=0}^n (-1)^{k} \cos\frac{k\pi}{2}
x^{n-k-1} \sigma_k^{(n)}   \nonumber\\
{\cal A}_{(n-1)(n-1)} & = & \sum_{l=0}^n (-1)^{l+1} \sin\frac{l\pi}{2}
x^{n-l} \sigma_l^{(n)} \hs\hs\hs. \nonumber
\end{eqnarray}
Developing the determinant of this matrix with respect to the last two columns
and taking into account eqs. (\ref{recfun}) and (\ref{loesung}), we obtain the
right hand side of equation (\ref{gleich1}), Q.E.D.

\subsubsection{The ``Inverse Yang-Lee" Point}

\,

A closed solution of the recursive equations (\ref{gleich1}) is also obtained
for
\EQ
B\left( 2 \sqrt{ \pi} \right) = { 2 \over 3} \,\,\, .
\EN
The reason is that, for this particular value of the coupling constant
the S-matrix of the Sinh-Gordon theory coincides with the inverse of the
$S$-matrix $S_{\rm YL}(\beta)$ of the Yang-Lee model \cite{Cardymus} or,
equivalently
\EQ
S(\beta,-\frac{2}{3})\,=\,S_{\rm YL}(\beta)\,\,\, .
\EN
Since the recursive equations (\ref{gleich1}) are invariant under
$B\rightarrow -B$ (see sect.\,4.2), a solution is provided by the same
combination of symmetric polynomials found for the Yang-Lee model
\cite{Smirnov2,YLZam}, i.e.
\EQ
P_n(x_1,x_2,\dots,x_n) \, = \,
{\rm det}\, {\cal B}(x_1,x_2,\ldots,x_n) \label{loesungyl}
\EN
with the following entries of the $(n-3) \times (n-3)$-matrix ${\cal B}$
\EQ
{\cal B}_{ij}\, = \, \sigma_{3j-2i+1} \,\,\, .
\EN
The proof is similar to the one of the previous section and exploits the
invariance of a determinant under linear combinations of the rows and the
columns. In this case the function $D_n$ is most conveniently expressed as
determinant of a $2 \times 2$-matrix
\EQ
D_n= \det \left( \begin{array}{cc}
\left(\sum\limits_{l=0}^n(-1)^{l}\cos\frac{l\pi}{3}x^{n-l}\sigma_l^{(n)}\right)&
\left(\sum\limits_{l=0}^n \cos\frac{l\pi}{3} x^{n-l} \sigma_l^{(n)} \right) \\
\left(\sum\limits_{l=0}^n(-1)^{l}\sin\frac{l\pi}{3}x^{n-l}\sigma_l^{(n)}\right)&
\left(\sum\limits_{l=0}^n \sin\frac{l\pi}{3} x^{n-l} \sigma_l^{(n)} \right) \\
\end{array}   \right)  \label{ddet}
\EN
Let us consider the $(n-1) \times (n-1)$-matrix entering the
expression $P_{n+2}(-x,x,x_1,\dots,x_n)$, i.e.
\EQ
{\cal B}_{ij} \, = \, \sigma_{3j-2i+1} - x^2 \sigma_{3j-2i-1},
\EN
By adding successively the $i^{th}$ row to row $(i-1)$ (starting with
$i=(n-1)$), we obtain
$$
{\cal B}_{ij} \, = \, \sigma_{3j-2i+1} - x^{2(n-i)} \sigma_{3j-2n+1}.
$$
Then by adding successively $x^6$ times the $i^{th}$ column to column $(i+2)$,
starting with $i=1$, the entries for the matrix ${\cal B}$ read
\EQ
{\cal B}_{ij} \, = \, \sum_{l=0} \sigma_{3j-2i-6l+1} x^{6l} - x^{2(n-i+3l)}
\sigma_{3j-2n-6l+1}. \label{ylentries}
\EN
Subtracting $x^6$ times of the row $(i+3)$ from row $i$ (starting with $i=1$)
we finally obtain the matrix
\EQ
{\cal B}^{(n-1) \times (n-1)} =
\left(
\begin{array}{ccc}
                             & 0                    &          0     \\
{\cal B}^{(n-3)\times (n-3)} & \vdots               &   \vdots       \\
                             &  0                   &      0         \\
   \ast \cdots \ast          & {\cal B}_{(n-2)(n-2)}&{\cal B}_{(n-2)(n-1)}\\
   \ast \cdots \ast          & {\cal B}_{(n-1)(n-2)}&{\cal B}_{(n-1)(n-1)} \\
\end{array}
\right)
\EN
where the entries of the $(2 \times 2)$ matrix in the lower right corner
are still given by (\ref{ylentries}). It is easy to prove that the
determinant of this $(2 \times 2)$ matrix in the lower right corner is equal
to (\ref{ddet}). Therefore, with the definition (\ref{loesungyl}), the
determinant of ${\cal B}^{(n-1) \times (n-1)}$ gives rise the right hand side
of (\ref{gleich1}). Q.E.D.

\section{Form Factors for Descendent Operators}

In this section we investigate the effect of the Lorentz transformation on the
space of solutions for all form factors denoted by ${\cal P}$.
This problem has been firstly addressed by Cardy and Mussardo
\cite{CMform} for the space of descendant operators of the Ising model.

The space of the form factors ${\cal P}$ can be decomposed as
\EQ
{\cal P} = \bigoplus_{s} \;\; {\cal P}_s  \quad ,
\EN
meaning that $F_n^{\cal O} \in {\cal P}_s$ if ${\cal O}$ has spin $s$.
On the rapidity variables a Lorentz transformation is realized by
$\be_i\rightarrow\be_i+\varepsilon$, i.e. $x_i\rightarrow e^{\varepsilon}x_i$.
Since the elementary symmetric polynomials are homogeneous functions of $x_i$,
under a Lorentz transformation they transform as
\EQ
\sigma_n(x_1,...,x_n) \rightarrow e^{ n \varepsilon} \sigma_n(x_1,...,x_n).
\EN
Hence, given a form factor $F_n^{\cal O}$ of an operator ${\cal O}$ with
spin $s$ which satisfies Watson's equations, a new function in
${\cal P}_{s+s'}$ which still satisfies Watson's equations can be defined by
\EQ
F_n^{\cal O'}(x_1,\dots ,x_n) = I_n^{s'}(x_1,\dots ,x_n) F_n^{\cal O}
(x_1,\dots ,x_n),
\label{descendant}
\EN
provided that $I_n^s$ is composed out of elementary symmetric polynomials.
Additional constraints on the $I_n^s$ are imposed by their invariance under
the kinematic residue equation
\EQ
I_{n+2}^s(-x,x,x_1,\dots,x_n) = I_n^s(x_1,\dots,x_n) \,\,\, .
\label{spinspace}
\EN
A basis in the space of solutions of eq.\,(\ref{spinspace}) is given by
symmetric polynomials $I_n^s$ satisfying the
recursion relations \cite{CMform}
\EQ
\sigma_{2k+1}^{(n)}\,=\,I_{n}^{2k+1}+\sigma_2^{(n)}\,I_{n}^{2k-1}+
\sigma_4^{(n)}\, I_{n}^{2k-3}+\ldots + \sigma_{2k}^{(n)}\,I_{n}^{1}
\,\,\, ,
\label{rec1}
\EN
where $s$ is equal to the spin of the conserved charges (\ref{conservedspin}).
A closed expression of $I_n^s$ as been obtained in \cite{Christe}
\EQ
I_n^{2s-1} = (-1)^{s+1} \det {\cal I}
\EN
where the entries of the $(s\times s)$-matrix ${\cal I}$
for $j=1,\ldots,s$ and $i=2,\ldots,s$ are
\EQ
{\cal I}_{1j} = \sigma_{2j-1} \qquad \qquad {\cal I}_{ij} = \sigma_{2j-2i+2}
\EN
i.e.
\EQ
{\cal I} =
\left(
\begin{array}{llllll}
\sigma_1 & \sigma_3 & \sigma_5 & \sigma_7 & \dots & \sigma_{2s-1}   \\
 1       & \sigma_2 & \sigma_4 & \sigma_6 & \dots & \sigma_{2s-2}   \\
 0       & 1        & \sigma_2 & \sigma_4 & \dots & \sigma_{2s-4}   \\
 0       & 0        &    1     & \sigma_2 & \dots & \sigma_{2s-2}   \\
 \vdots  & \vdots   &  \vdots  & \vdots   & \ddots & \vdots         \\
\end{array}
\right)
\EN
The determinant of ${\cal I}$ will always be of order $2s-1$ as required.

As it was first noticed in \cite{Cardymus}, eqs.\,(\ref{descendant})
naturally provides a grading in the space of matrix elements of local
operators in an integrable massive field theory. In fact, given an invariant
polynomial $I_n^s$, eq.\,(\ref{descendant}) defines form factors of an operator
${\cal O}_s'$ which, borrowing the terminology of Conformal Field
Theories \cite{BPZ}, is natural to call {\em descendant operator} of the
spinless field ${\cal O}$. In particular, choosing ${\cal O}$ to be the trace
of the stress-energy tensor, the form factors defined by
eq.\,(\ref{descendant}) are related to the matrix elements of the higher
conserved currents, as can be easily seen by eq.\,(\ref{charges}) and by the
fact that the symmetric polynomials which appear as eigenvalues of the
conserved charges ${\cal Q}_s$
\EQ
s_k\,=\, x_1^k+x_2^k+\ldots + x_n^k\,\, ,
\EN
can be expressed in terms of the invariant polynomials $I_n^s$. Indeed they
satisfy the recursive relation
\EQ
s_k-s_{k-1}\sigma_1+s_{k-2}\sigma_2-\ldots
+(-1)^{k-1}s_1\sigma_{k-1}+(-1)^k k\sigma_k\,=\,0 \,\, ,
\label{rec2}
\EN
that, together with eq.\,(\ref{rec1}), permits to express $s_k$ in terms
of the invariant polynomials $I_n^{s}$.

\resection{Form factors and $c$-theorem}

\,

As mentioned in sect.\,3, the Sinh-Gordon model can be regarded as
deformation of the free massless theory with central charge $c=1$.
This fixed point governs the ultraviolet behaviour of the model whereas
the infrared behaviour corresponds to a massive field theory with
central charge $c=0$. Going from the short- to large-distances, the
variation of the central charge is dictated by the $c$-theorem of
Zamolodchikov \cite{cth}. An integral version of this theorem has
been derived by Cardy \cite{Cardycth} and related to the spectral
representation of the two-point function of the trace of the stress-energy
tensor in \cite{Friedan,Freedman}, i.e.
\EQ
\Delta c \,=\, \int_0^{\infty} d\mu\, c_1(\mu)\,\, ,
\label{variation}
\EN
where $c_1(\mu)$ is given by
\begin{eqnarray}
& & c_1(\mu)\,=\,\frac{6}{\pi^2}\frac{1}{\mu^3} {\rm Im}\, G(p^2=-\mu^2)
\,\, ,
\label{spectral} \\
& & G(p^2) \,=\, \int d^2 x\, e^{-ip\dot x} \,
<0|\Theta(x)\Theta(0)|0>_{\rm conn}
\nonumber \,\,\, .
\end{eqnarray}
Inserting a complete set of in-state into (\ref{spectral}), we can express
the function $c_1(\mu)$ in terms of the form factors $F_{2n}^{\Theta}$
\begin{eqnarray}
c_1(\mu)& =&\frac{12}{\mu^3} \sum_{n=1}^{\infty} \frac{1}{(2n)!}
\int\frac{d\beta_1\ldots d\beta_{2n}}{(2\pi)^{2n}}\,
\mid F_{2n}^{\Theta}(\beta_1,\ldots, \beta_{2n})\mid^2 \\
& & \,\,\, \times \,
\delta(\sum_i m\sinh\beta_i)\,\delta(\sum_i m\cosh\beta_i-\mu)\,\,\, .
\nonumber
\end{eqnarray}
For the Sinh-Gordon theory $\Delta c=1$ and it is interesting to study
the convergence of this series increasing the number of intermediate
particles. For the two-particle contribution, we have the following
expression
\EQ
\Delta c^{(2)}\,=\,\frac{3}{2 F^2_{\rm min}(i\pi)}
\,\int_{0}^{\infty} \frac{d\beta}{\cosh^4\beta} \,|F_{\rm min}(2\beta)|^2
\,\,\, .
\EN
The numerical results for different values of the coupling
constant $g^2/4\pi$ are listed in Table 1. It is evident that the sum rule
is saturated by the two-particle form factor also for large values of the
coupling constant.  Hence, the expansion in the number of intermediate
particles results in a fast convergent series, as it is confirmed by the
computation of the next terms involving the form factor with four and
six particles.

\resection{Conclusions}

\,

The computation of the Green functions is a central problem in a Quantum Field
Theory. For integrable models, a promising approach to this question is
given by the bootstrap principle applied to the computation of the matrix
elements of local operators. In this paper we have investigated the form
factors of the most representative fields of the $Z_2$ sectors of the
Sinh-Gordon model, i.e. the field $\phi(x)$ and the trace of the stress-energy
tensor $\Theta(x)$. The simplicity of the Sinh-Gordon model permits
clarification of the basic properties of the local operators and their matrix
elements in a QFT, without being masked by algebraic complexities due to the
structure of the bound states. Compared to the usual method of
computing correlation functions in terms of a perturbative series in the
coupling constant, the form factor approach is extremely advantageous
for two reasons. Firstly, the coupling constant dependence of the
correlation functions is encoded (to all orders in $g$) into the expression
of $F_{\rm min}(b,B)$, eq.\,(\ref{integral}), and into the solutions of the
recursive equations (\ref{recursive}) for the pre-factors $K_n$ entering the
form factors, eq.\,(\ref{parametrization}). Secondly, even for not large
values of the distances, the resulting expressions of the correlation
functions as an infinite series over the multi-particle form factors are
actually dominated by the lowest number of particle terms and therefore
present a very fast rate of convergence.

\vspace{1cm}

\section*{Acknowledgments}

A.F. and G.M. are grateful for the hospitality at the International School
for Advanced Studies and Imperial College, respectively. We are grateful to
S. Cecotti, F. Colomo, S. Elitzur, M. Nolasco, D. Olive, I. Pesando and
A. Schwimmer for useful discussions.

\newpage

\hs

\vspace{25mm}

{\bf Figure Captions}

\vspace{1cm}

\begin{description}
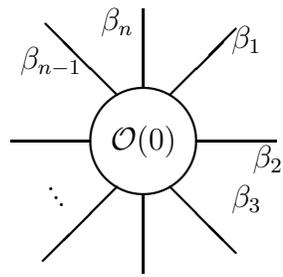
\item [Figure 1]. Form factors of the operator ${\cal O}(0)$.
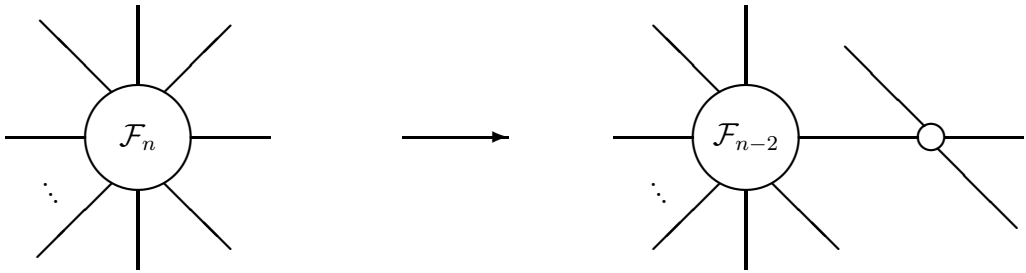
\item [Figure 2]. Kinematical recursive equation for the form factor
${\cal F}_n$
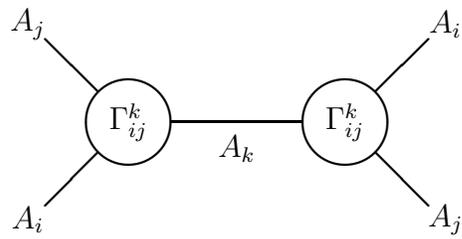
\item [Figure 3]. Bound state pole in scattering amplitude.
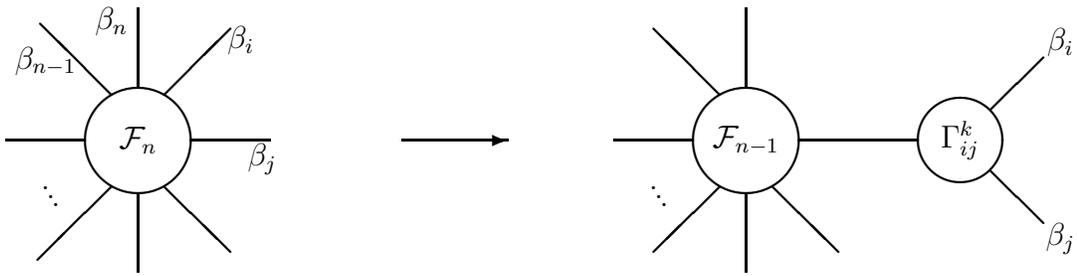
\item [Figure 4]. Bound state recursive equation for the form factor ${\cal
F}_n$.
\item [Figure 5]. Graphs of $|F_{\rm min}(\beta,B)/{\cal N}|^2$ as function
of $\beta$ for different values of $B(g)$.
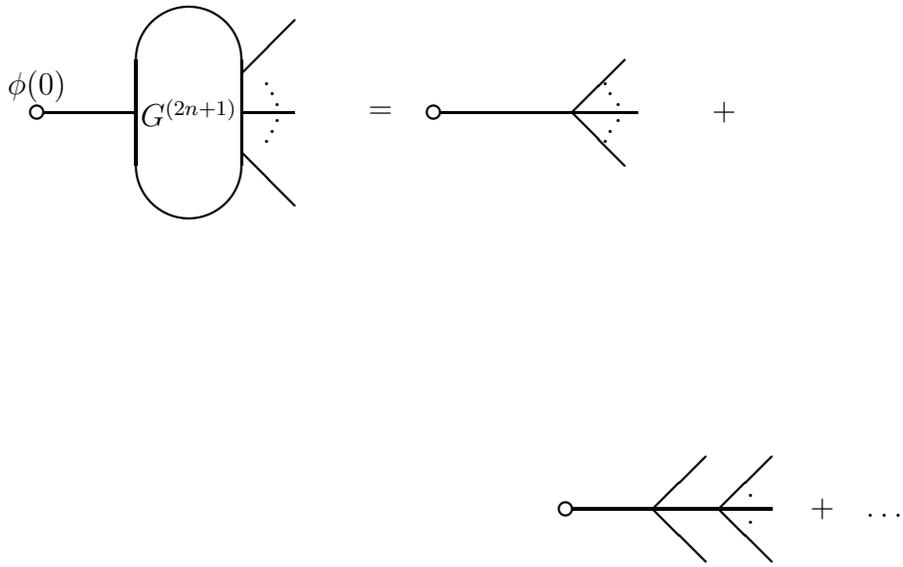
\item [Figure 6]. Lowest terms in the perturbative expression of the
form factors of the elementary field $\phi(0)$. $G^{(2n+1)}$ is the
Green function with $2n+1$ external legs.
\end{description}

\newpage

\hs

\vspace{25mm}

{\bf Table Caption}

\vspace{1cm}

\begin{description}
\item [Table 1]. The first two-particle term entering the sum rule of the
$c$-theorem.
\end{description}

\newpage

%Begin of Figures

\newpage

\hs

\vspace{3cm}

\begin{center}
\begin{picture}(280,260)
\thicklines
\put(120,180){\circle{40}}
\put(100,180){\line(-1,0){30}}
\put(120,200){\line(0,1){30}}
\put(120,160){\line(0,-1){30}}
\put(120,180){\makebox(0,0){${\cal O}(0)$}}
\put(110,197.33){\line(-1,1){27}}
\put(159,218){\makebox(0,0){$\beta_1$}}
\put(159,158){\makebox(0,0){$\beta_3$}}
\put(110,162.67){\line(-1,-1){28}}
\put(110,225){\makebox(0,0){$\beta_n$}}
\put(85,210){\makebox(0,0){$\beta_{n-1}$}}
\put(140,180){\line(1,0){30}}
\put(85,162){\makebox(0,0){.}}
\put(87,158.55){\makebox(0,0){.}}
\put(89,155.75){\makebox(0,0){.}}
\put(167,173){\makebox(0,0){$\beta_2$}}
\put(130,197.33){\line(1,1){25}}
\put(130,162.67){\line(1,-1){25}}
\end{picture}
\end{center}
\vspace{25mm}
\begin{center}
{\bf Figure 1}
\end{center}

\newpage

\newpage

\begin{center}
\begin{picture}(280,260)
\thicklines
\put(-30,0){\circle{40}}
\put(-50,0){\line(-1,0){30}}
\put(-30,20){\line(0,1){30}}
\put(-30,-20){\line(0,-1){30}}
\put(-30,0){\makebox(0,0){${\cal F}_n$}}
\put(-40,17.33){\line(-1,1){27}}
%\put(9,38){\makebox(0,0){$\beta_1$}}
%\put(9,-38){\makebox(0,0){$\beta_3$}}
\put(-40,-17.33){\line(-1,-1){28}}
%\put(-40,45){\makebox(0,0){$\beta_n$}}
%\put(-65,30){\makebox(0,0){$\beta_{n-1}$}}
\put(-10,0){\line(1,0){30}}
\put(-65,-18){\makebox(0,0){.}}
\put(-63,-21.45){\makebox(0,0){.}}
\put(-61,-24.25){\makebox(0,0){.}}
%\put(17,-7){\makebox(0,0){$\beta_2$}}
\put(-20,17.33){\line(1,1){25}}
\put(-20,-17.33){\line(1,-1){25}}
\put(70,0){\vector(1,0){40}}
\put(200,0){\circle{40}}
\put(200,0){\makebox(0,0){${\cal F}_{n-2}$}}
\put(210,-17.33){\line(1,-1){25}}
\put(200,20){\line(0,1){30}}
\put(200,-20){\line(0,-1){30}}
\put(180,0){\line(-1,0){30}}
\put(165,-18){\makebox(0,0){.}}
\put(167,-21.45){\makebox(0,0){.}}
\put(169,-24.25){\makebox(0,0){.}}
\put(190,17.33){\line(-1,1){25}}
\put(190,-17.33){\line(-1,-1){25}}
\put(220,0){\line(1,0){45}}
\put(270,0){\circle{10}}
\put(275,0){\line(1,0){30}}
%\put(303,-7){\makebox(0,0){$\beta_2$}}
\put(267.5,4.331){\line(-1,1){30}}
%\put(300,39){\makebox(0,0){$\beta_1$}}
%\put(300,-39){\makebox(0,0){$\beta_3$}}
\put(272.5,-4.331){\line(1,-1){30}}
\end{picture}
\end{center}
\vspace{25mm}
\begin{center}
{\bf Figure 2}
\end{center}

\newpage

\hs

\vspace{3cm}

\hspace{10cm}

\begin{center}
\begin{picture}(300,160)
\thicklines
\put(80,70){\circle{32}}
\put(80,70){\makebox(0,0){$\Gamma_{ij}^k$}}
\put(96,70){\line(1,0){50}}
\put(68.5,58.5){\line(-1,-1){20}}
\put(68.5,81.5){\line(-1,1){20}}
\put(162,70){\circle{32}}
\put(162,70){\makebox(0,0){$\Gamma_{ij}^k$}}
\put(173.5,81.5){\line(1,1){20}}
\put(173.5,58.5){\line(1,-1){20}}
\put(200,107){\makebox(0,0){$A_i$}}
\put(200,33){\makebox(0,0){$A_j$}}
\put(42,33){\makebox(0,0){$A_i$}}
\put(42,107){\makebox(0,0){$A_j$}}
\put(121,60){\makebox(0,0){$A_k$}}
\end{picture}
\end{center}
\vspace{5mm}
\begin{center}
{\bf Figure 3}
\end{center}

\newpage

\begin{center}
\begin{picture}(280,260)
\thicklines
\put(-30,0){\circle{40}}
\put(-50,0){\line(-1,0){30}}
\put(-30,20){\line(0,1){30}}
\put(-30,-20){\line(0,-1){30}}
\put(-30,0){\makebox(0,0){${\cal F}_n$}}
\put(-40,17.33){\line(-1,1){27}}
\put(9,38){\makebox(0,0){$\beta_i$}}
%\put(9,-38){\makebox(0,0){$\beta_3$}}
\put(-40,-17.33){\line(-1,-1){28}}
\put(-40,45){\makebox(0,0){$\beta_n$}}
\put(-65,30){\makebox(0,0){$\beta_{n-1}$}}
\put(-10,0){\line(1,0){30}}
\put(-65,-18){\makebox(0,0){.}}
\put(-63,-21.45){\makebox(0,0){.}}
\put(-61,-24.25){\makebox(0,0){.}}
\put(17,-7){\makebox(0,0){$\beta_j$}}
\put(-20,17.33){\line(1,1){25}}
\put(-20,-17.33){\line(1,-1){25}}
\put(70,0){\vector(1,0){40}}
\put(200,0){\circle{40}}
\put(200,0){\makebox(0,0){${\cal F}_{n-1}$}}
\put(210,-17.33){\line(1,-1){25}}
\put(200,20){\line(0,1){30}}
\put(200,-20){\line(0,-1){30}}
\put(180,0){\line(-1,0){30}}
\put(165,-18){\makebox(0,0){.}}
\put(167,-21.45){\makebox(0,0){.}}
\put(169,-24.25){\makebox(0,0){.}}
\put(190,17.33){\line(-1,1){25}}
\put(190,-17.33){\line(-1,-1){25}}
\put(220,0){\line(1,0){45}}
\put(281,0){\circle{32}}
\put(281,0){\makebox(0,0){$\Gamma_{ij}^k$}}
\put(292.5,11.5){\line(1,1){20}}
\put(292.5,-11.5){\line(1,-1){20}}
\put(319,37){\makebox(0,0){$\beta_i$}}
\put(319,-37){\makebox(0,0){$\beta_j$}}
%\put(270,0){\circle{10}}
%\put(275,0){\line(1,0){30}}
%\put(303,-7){\makebox(0,0){$\beta_j$}}
%\put(272.5,4.331){\line(1,1){30}}
%\put(300,39){\makebox(0,0){$\beta_i$}}
%\put(300,-39){\makebox(0,0){$\beta_j$}}
%\put(272.5,-4.331){\line(1,-1){30}}
\end{picture}
\end{center}
\vspace{25mm}
\begin{center}
{\bf Figure 4}
\end{center}

\newpage

\hs

\begin{center}
\begin{picture}(400,350)
\thicklines
\put(50,220){\circle{5}}
\put(50,230){\makebox(0,0){$\phi(0)$}}
\put(52.5,220){\line(1,0){35}}
\put(107.5,220){\oval(40,80)}
\put(107.5,220){\makebox(0,0){$G^{(2n+1)}$}}
\put(127.5,235){\line(1,1){20}}
\put(137,231){\makebox(0,0){$.$}}
\put(139,227){\makebox(0,0){$.$}}
\put(141,223){\makebox(0,0){$.$}}
\put(127.5,220){\line(1,0){20}}
\put(141,217){\makebox(0,0){$.$}}
\put(139,213){\makebox(0,0){$.$}}
\put(137,209){\makebox(0,0){$.$}}
\put(127.5,205){\line(1,-1){20}}
\put(180,220){\makebox(0,0){$=$}}
\put(200,220){\circle{5}}
\put(202.5,220){\line(1,0){50}}
\put(252.5,220){\line(1,1){20}}
\put(252.5,220){\line(1,0){25}}
\put(252.5,220){\line(1,-1){20}}
\put(270,223){\makebox(0,0){$.$}}
\put(267,227){\makebox(0,0){$.$}}
\put(265,231){\makebox(0,0){$.$}}
\put(270,217){\makebox(0,0){$.$}}
\put(267,213){\makebox(0,0){$.$}}
\put(265,209){\makebox(0,0){$.$}}
\put(310,220){\makebox(0,0){$+$}}
\put(250,70){\circle{5}}
\put(252.5,70){\line(1,0){30.5}}
\put(283,70){\line(1,0){25}}
\put(283,70){\line(1,1){20}}
\put(283,70){\line(1,-1){20}}
\put(308,70){\line(1,0){20}}
\put(308,70){\line(1,1){20}}
\put(308,70){\line(1,-1){20}}
\put(320,75){\makebox(0,0){$.$}}
\put(320,65){\makebox(0,0){$.$}}
\put(360,70){\makebox(0,0){$+\,\,\,\,\,\ldots$}}
\end{picture}
\end{center}
\vspace{25mm}
\begin{center}
{\bf Figure 6}
\end{center}

%End of figures

%%%%%%%%%%%%%%%%%%%%%%%%%%%%%%%%%%%%%%%%%%%%%%%%%%%%%%%%%%%%%%%%%%%%%%%%%%%%%

\newpage

%Beginning of tables

\begin{table}

$$ \begin{array}{|c|c|c|}
 \hline
& &  \\
B
& \frac{g^2}{4\pi}
& \Delta\,c^{(2)}
\\
   &        &        \\
\hline
   &        &        \\
\frac{1}{500} & \frac{2}{999} & 0.9999995 \\
\frac{1}{100} & \frac{2}{199} & 0.9999878 \\
\frac{1}{10} & \frac{2}{19} &   0.9989538  \\
\frac{3}{10} & \frac{6}{17} &   0.9931954  \\
\frac{2}{5} & \frac{1}{2} &     0.9897087  \\
\frac{1}{2} & \frac{2}{3} &     0.9863354  \\
\frac{2}{3} & 1 &     0.9815944  \\
\frac{7}{10} & \frac{14}{13} &     0.9808312  \\
\frac{4}{5} & \frac{4}{3} &     0.9789824  \\
1 & 2 &     0.9774634  \\
& & \\ \hline
\end{array}
$$
\end{table}
\begin{center}
{\bf Table 1}
\end{center}

% End tables

\end{document}